\documentclass[11pt,onecolumn]{article}
\usepackage[dvips]{graphics}
\usepackage[dvips]{color}
\usepackage{amssymb}
\usepackage{amsmath}
\usepackage{amsthm}
\usepackage[dvips]{geometry}
\usepackage{setspace}

\theoremstyle{plain}
\newtheorem{theorem}{Theorem}

\newtheorem*{conjecture}{Conjecture}

\newcommand{\half}{\frac{1}{2}}

\begin{document}

\title{Capacity of a Class of Deterministic Relay Channels}

\author{Thomas M. Cover and Young-Han Kim%
\thanks{Email: cover@stanford.edu, yhk@ucsd.edu}}
\date{\vspace*{-2em}}
\maketitle

\begin{abstract}
The capacity of a class of deterministic relay channels with the
transmitter input $X$, the receiver output $Y$, the relay output $Y_1
= f(X, Y)$, and a separate communication link from the relay to the
receiver with capacity $R_0$, is shown to be
\[
C(R_0) = \max_{p(x)} \min \{ I(X;Y) + R_0,\; I(X;Y, Y_1) \}.
\]
Thus every bit from the relay is worth exactly one bit to the
receiver.  Two alternative coding schemes are presented that achieve
this capacity.  The first scheme, ``hash-and-forward'', is based on a
simple yet novel use of random binning on the space of relay outputs,
while the second scheme uses the usual ``compress-and-forward''.  In
fact, these two schemes can be combined together to give a class of
optimal coding schemes.  As a corollary, this relay capacity result
confirms a conjecture by Ahlswede and Han on the capacity of a channel
with rate-limited state information at the decoder in the special case
when the channel state is recoverable from the channel input and the
output.
\end{abstract}

\section{Introduction with Gaussian Relay}
\label{sec:intro}
\begin{figure}[h!]
\vspace*{12pt}
\begin{center}
\input{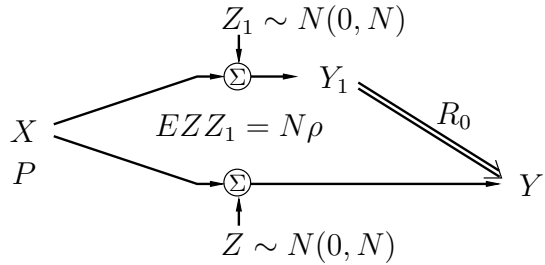}
\end{center}
\caption{Gaussian relay channel with a noiseless link.}
\label{fig:gaussian}
\end{figure}

Consider the Gaussian relay problem shown in
Figure~\ref{fig:gaussian}.  
Suppose the receiver $Y$ and the relay
$Y_1$ each receive information about the transmitted signal $X$ of
power $P$. Specifically, let 
\begin{align*}
Y&= X + Z\\
Y_1 &= X + Z_1,
\end{align*}
where $(Z, Z_1)$ have correlation coefficient $\rho$ and are jointly
Gaussian with zero mean and equal variance $EZ^2 = EZ_1^2 = N$. What
should the relay $Y_1$ say to the ultimate receiver $Y$? If the relay
sends information at rate $R_0$, what is the capacity $C(R_0)$ of the
resulting relay channel?

We first note that the capacity from $X$ to $Y$, ignoring the relay,
is
\[
C(0) =\half \log\left(1+\frac{P}{N}\right) \quad\text{bits per transmission.}
\]
The channel from the relay $Y_1$ to the ultimate receiver $Y$ has
capacity $R_0$.  This relay information is sent on a side channel that
does not affect the distribution of $Y$, and the information becomes
freely available to $Y$ as long as it doesn't exceed rate $R_0$. We
focus on three cases for the noise correlation $\rho$: $\rho= 1, 0,$
and $-1$.

If $\rho=1$, then $Y_1=Y$, the relay is useless, and the capacity of
the relay channel is $C(R_0)=(1/2)\log(1+P/N) = C(0)$ for all $R_0 \ge
0$.

Now consider $\rho=0$, i.e., the noises $Z$ and $Z_1$ are
independent. Then the relay $Y_1$ has no more information about $X$
than does $Y$, but the relay furnishes an independent look at
$X$. What should the relay say to $Y$?  This capacity $C(R_0)$,
mentioned in \cite{Cover1987b}, remains unsolved and typifies the
primary open problem of the relay channel.  As a partial converse,
Zhang~\cite{Zhang1988} obtained the strict inequality $C(R_0) < C(0) +
R_0$ for all $R_0 > 0$.

\begin{figure}[h]
\vspace*{12pt}
\begin{center}
\input{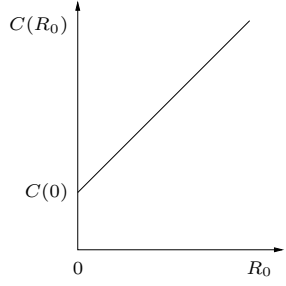}
\end{center}
\caption{Gaussian relay capacity $C(R_0)$ vs.\@ the relay
information rate $R_0$.}
\label{fig:graph}
\end{figure}

How about the case $\rho=-1$? This is the problem that we solve and
generalize in this note.  Here the relay, while having no more
information than the receiver $Y$, has much to say, since knowledge of
$Y$ and $Y_1$ allows the perfect determination of $X$. However, the
relay is limited to communication at rate $R_0$. Thus, by a simple
cut-set argument, the total received information is limited to $C(0) +
R_0$ bits per transmission. We argue that this rate can actually be
achieved. Since it is obviously the best possible rate, the capacity
for $\rho=-1$ is given as
\[
C(R_0)= C(0) + R_0.
\]
(See Figure~\ref{fig:graph}.)
Every bit sent by the relay counts as one bit of information, despite
the fact that the relay doesn't know what it is doing.

We present two distinct methods of achieving the capacity. Our first
coding scheme consists of hashing $Y_1^n$ into $nR_0$ bits, then
checking the $2^{nC(R_0)}$ codewords $X^n(W)$, $W \in 2^{nC(R_0)}$,
one by one, with respect to the ultimate receiver's output $Y^n$ and
the hash check of $Y_1^n$.  More specifically, we check whether the
corresponding estimated noise $\hat{Z}^n=Y^n-X^n(W)$ is typical, and
then check whether the resulting $Y_1^n(W) =X^n(W) + \hat{Z}^n$
satisfies the hash of the observed $Y_1^n$.  Since the typicality
check reduces the uncertainty in $X^n(W)$ by a factor of $2^{n C(0)}$
while the hash check reduces the uncertainty by a factor of $2^{n
R_0}$, we can achieve the capacity $C(R_0) = C(0) + R_0$.

It turns out hashing is not the unique way of achieving $C(R_0) = C(0)
+ R_0$.  We can compress $Y_1^n$ into $\hat{Y}_1^n$ using $n R_0$ bits
with $Y^n$ as side information in the same manner as in Wyner--Ziv
source coding~\cite{Wyner--Ziv1976}, which requires
\[
R_0 = I(Y_1; \hat{Y}_1 | Y).
\]
Thus, $nR_0$ bits are sufficient to reveal $\hat{Y}_1^n$ to the
ultimate receiver $Y^n$.  Then, based upon the observation $(Y^n,
\hat{Y}_1^n)$, the decoder can distinguish $2^{nR}$ messages if
\[
R < R^* := I(X; Y, \hat{Y}_1).
\]

For this scheme, we now choose the appropriate distribution of
$\hat{Y}_1$ given $Y_1$.  Letting
\[
\hat{Y}_1 = Y_1 + U,
\]
where $U \sim N(0, \sigma^2)$ is independent of $(X, Z, Z_1)$, we
can obtain the following parametric expression of $R^*(R_0)$
over all $\sigma^2 > 0$:
\begin{align}
\label{eq:param1}
R^*(\sigma^2) &= I(X; Y, \hat{Y}_1) =
\frac{1}{2} \log 
\left(
\frac{
(P+N)\sigma^2 + 4 P N }{N\sigma^2}
\right)\\
\label{eq:param2}
R_0(\sigma^2) &= I(Y_1; \hat{Y}_1|Y)
= \frac{1}{2} \log
\left(
\frac{
(P+N)\sigma^2 + 4 PN}{(P+N)\sigma^2}
\right).
\end{align}
Setting $R_0(\sigma_0^2) = R_0$ in \eqref{eq:param2}, solving for
$\sigma_0^2$, and inserting it in \eqref{eq:param1}, we find the
achievable rate is given by
\[
R^*(\sigma_0^2) = R_0 + \frac{1}{2}
\log \left(1 + \frac{P}{N}\right) = C(0) + R_0,
\]
so ``compress-and-forward'' also achieves the capacity.

Inspecting what it is about this problem that allows this solution, we
see that the critical ingredient is that the relay output $Y_1=f(X,Y)$
is a deterministic function of the input $X$ and the receiver output
$Y$.  This leads to the more general result stated in
Theorem~\ref{thm:main} in the next section.

\section{Main Result}
\label{sec:main}

\begin{figure}[ht]
\begin{center}
\input{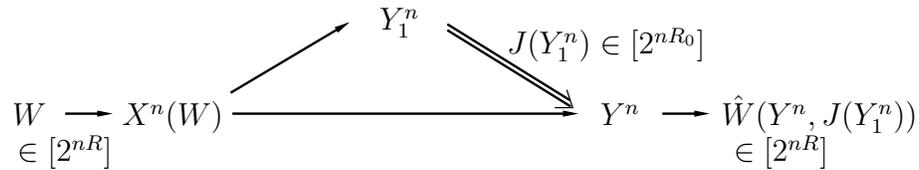}
\end{center}
\caption{Relay channel with a noiseless link.}
\label{fig:det-relay}
\end{figure}

We consider the following relay channel with a noiseless link as
depicted in Figure~\ref{fig:det-relay}.  
We define a \emph{relay
channel with a noiseless link} $(\mathcal{X}, p(y, y_1|x),
\mathcal{Y}\times \mathcal{Y}_1, R_0)$ as the channel where the input
signal $X$ is received by the relay $Y_1$ and the receiver $Y$ through
a channel $p(y,y_1|x)$, and the relay can communicate to the receiver
over a separate noiseless link of rate $R_0$.  We wish to communicate
a message index $W \in [2^{nR}] = \{1,2,\ldots, 2^{nR}\}$ reliably
over this relay channel with a noiseless link.%
\footnote{Henceforth, the notation $i \in [2^{nR}]$ is interpreted to
mean $i \in \{1,2,\ldots, 2^{nR}\}$.}  We specify a $(2^{nR}, n)$ code
with an encoding function $X^n: [2^{nR}] \to \mathcal{X}^n$, a relay
function $J: \mathcal{Y}_1^n \to [2^{nR_0}]$, and the decoding
function $\hat{W}: \mathcal{Y}^n \times [2^{nR_0}] \to [2^{nR}]$.  The
probability of error is defined by $P_e^{(n)} = \Pr \{ W \ne
\hat{W}(Y^n,J(Y_1^n) \}, $ with the message $W$ distributed uniformly
over $[2^{nR}]$.  The capacity $C(R_0)$ is the supremum of the rates
$R$ for which $P_e^{(n)}$ can be made to tend to zero as $n \to
\infty$.

We state our main result.

\begin{theorem}
For the relay channel $(\mathcal{X}, p(y,y_1|x), \mathcal{Y} \times
\mathcal{Y}_1)$ with a noiseless link of rate $R_0$ from the relay to
the receiver, if the relay output $Y_1 = f(X,Y)$ is a deterministic
function of the input $X$ and the receiver output $Y$, then the
capacity is given by
\[
C(R_0) = \max_{p(x)} \min \{ I(X;Y) + R_0,\; I(X;Y_1, Y) \}.
\]
\label{thm:main}
\end{theorem}

The converse is immediate from the simple application of the max-flow
min-cut theorem on information flow~\cite[Section
15.10]{Cover--Thomas2006}.

The achievability has several interesting features.  First, as we will
show in the next section, a novel application of random binning
achieves the cut-set bound.  In this coding scheme, the relay simply
sends the hash index of its received output $Y_1^n$.

What is perhaps more interesting is that the same capacity can be
achieved also via the well-known ``compress-and-forward'' coding
scheme of Cover and El Gamal~\cite{Cover--El-Gamal1979}.  In this
coding scheme, the relay compresses its received output $Y_1^n$ as in
Wyner--Ziv source coding with the ultimate receiver output $Y^n$ as
side information.

In both coding schemes, every bit of relay information carries one bit
of information about the channel input, although the relay does not
know the channel input.  And the relay information can be summarized
in a manner completely independent of geometry (random binning) or
completely dependent on geometry (random covering).

More surprisingly, we can partition the relay space using both random
binning and random covering.  Thus, a combination of
``hash-and-forward'' and ``compress-and-forward'' achieves the
capacity.

The next section proves the achievability using the
``hash-and-forward'' coding scheme.  The ``compress-and-forward''
scheme is deferred to Section~\ref{sec:second} and the combination
will be discussed in Sections~\ref{sec:discuss} and \ref{sec:third}.

\section{Proof of Achievability (Hash and Forward)}
\label{sec:first}
We combine the usual random codebook generation with list decoding and
random binning of the relay output sequences:

\emph{Codebook generation.}  Generate $2^{nR}$ independent codewords
$X^n(w)$ of length $n$ according to $\prod_{i=1}^n p(x_i)$.
Independently, assign all possible relay output sequences in
$|\mathcal{Y}_1|^n$ into $2^{nR_0}$ bins uniformly at random.

\emph{Encoding.} To send the message index $w \in [2^{nR}]$, the
transmitter sends the codeword $X^n(w)$.  Upon receiving the output
sequence $Y_1^n$, the relay sends the bin index $b(Y_1^n)$ to the
receiver.

\emph{Decoding.} Let $A_\epsilon^{(n)}$~\cite[Section
7.6]{Cover--Thomas2006} denote the set of jointly typical sequences
$(x^n, y^n) \in \mathcal{X}^n \times \mathcal{Y}^n$ under the
distribution $p(x,y)$.  The receiver constructs a list
\[
L(Y^n) = \{X^n(w) : w \in [2^{nR}], (X^n(w), Y^n) \in
A_{\epsilon}^{(n)}\}
\]
of codewords $X^n(w)$ that are jointly typical with $Y^n$.  Since the
relay output $Y_1$ is a deterministic function of $(X, Y)$, then for
each codeword $X^n(w)$ in $L(Y^n)$, we can determine the corresponding
relay output $Y_1^n(w) = f(X^n(w), Y^n)$ exactly.  The receiver
declares $\hat{w} = w$ was sent if there exists a unique codeword
$X^n(w)$ with the corresponding relay bin index $b(f(X^n(w), Y^n))$
matching the true bin index $b(Y_1^n)$ received from the relay.

\begin{figure}[h]
\begin{center}
\input{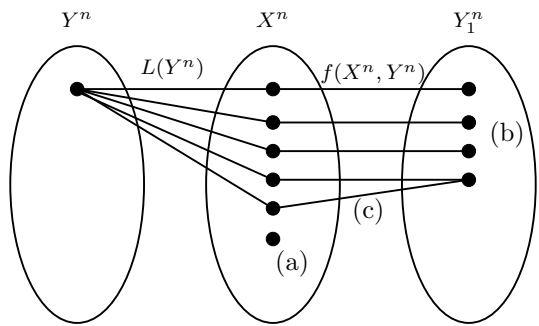}
\end{center}
\caption{Schematic diagram of ``hash-and-forward'' coding scheme.
The error happens when (a) the true codeword is not jointly typical
with $Y^n$, (b) there is more than one $Y_1^n$ for the same bin index,
or
(c) there is more than one $X^n$ jointly typical with
$(Y^n, Y_1^n)$.}
\label{fig:scheme1}
\end{figure}

\emph{Analysis of the probability of error.}  Without loss of
generality, assume $W=1$ was sent.  The sources of error are as
follows (see Figure~\ref{fig:scheme1}):
\begin{enumerate}
\item[(a)] The pair $(X^n(1), Y^n)$ is not typical.  The probability
of this event vanishes as $n$ tends to infinity.

\item[(b)] The pair $(X^n(1), Y^n)$ is typical, but there is more than
one relay output sequence $Y_1^n(w) = f(X^n(w), Y^n)$ with the
observed bin index, i.e., $b(Y_1^n(1)) = b(Y_1^n(w))$.  By Markov's
inequality, the probability of this event is upper bounded by the
expected number of codewords in $L(Y^n)$ with the corresponding relay
bin index equal to the true bin index $b(Y_1^n(1))$.  Since the bin
index is assigned independently and uniformly, this is bounded by
\[
2^{nR}\, 2^{-n (I(X;Y)-\epsilon)}\, 2^{-nR_0},
\]
which vanishes asymptotically as $n \to \infty$ if $R < I(X;Y) + R_0 -
\epsilon$.

\item[(c)] The pair $(X^n(1), Y^n)$ is typical and there is exactly
one $Y_1^n(w)$ matching the true relay bin index, but there is more
than one codeword $X^n(w)$ that is jointly typical with $Y^n$ and
corresponds to the same relay output $Y_1^n$, i.e., $f(X^n(1),Y^n) =
f(X^n(w), Y^n)$.  The probability of this kind of error is upper
bounded by
\[
2^{nR} \, 2^{- n(I(X; Y, Y_1) - \epsilon)},
\]
which vanishes asymptotically if $R < I(X; Y, Y_1) - \epsilon$.\qedhere
\end{enumerate}

\vspace*{18pt}

\section{Related Work}
\label{sec:related}

The general relay channel was introduced by van der
Meulen~\cite{van-der-Meulen1971}.  We refer the readers to Cover and
El Gamal~\cite{Cover--El-Gamal1979} for the history and the definition
of the general relay channel.  For recent progress, refer to Kramer
et al.~\cite{Kramer--Gastpar--Gupta2005}, El Gamal et
al.~\cite{El-Gamal--Hassanpour--Mammen2006}, and the references
therein.

We recall the following achievable rate for the general relay channel
investigated in~\cite{Cover--El-Gamal1979}.

\begin{theorem}[{\cite[Theorem 7]{Cover--El-Gamal1979}}]
For any relay channel $(\mathcal{X}\times\mathcal{X}_1, p(y, y_1| x,
x, x_1), \mathcal{Y}\times\mathcal{Y}_1)$, the capacity $C$ is lower
bounded by
\[
C \ge \sup \min \{I(X; Y, \hat{Y}_1| X_1, U) + 
I (U; Y_1|X_1, V),
\; I(X, X_1; Y) - I(\hat{Y}_1; Y_1| X, X_1, Y, U) \}
\]
where the supremum is taken over all joint probability distributions
of the form
\[
p(u,v,x,x_1,y,y_1,\hat{y}_1) = p(v) p(u|v) p(x|u)
p(x_2|v) p(y,y_1|x,x_1) p(\hat{y}_1|x_1, y_1, u) 
\]
subject to the constraint
\[
I({Y}_1; \hat{Y}_1 | X_1, Y, U) \le I(X_1; Y | V).
\]
\label{thm:ceg}
\end{theorem}

Roughly speaking, the achievability of the rate in
Theorem~\ref{thm:ceg} is based on a superposition of
``decode-and-forward'' (in which the relay decodes the message and
sends it to the receiver) and ``compress-and-forward'' (in which the
relay compresses its own received signal without decoding and sends it
to the receiver).  This coding scheme turns out to be optimal for many
special cases; Theorem~\ref{thm:ceg} reduces to the capacity when the
relay channel is degraded or reversely
degraded~\cite{Cover--El-Gamal1979} and when there is feedback from
the receiver to the relay~\cite{Cover--El-Gamal1979}.

Furthermore, for the semideterministic relay channel with the sender
$X$, the relay sender $X_1$, the relay receiver $Y_1 = f(X, X_1),$ and
the receiver $Y$, El Gamal and Aref~\cite{El-Gamal--Aref1982} showed
that Theorem~\ref{thm:ceg} reduces to the capacity given by
\begin{equation}
C = \max_{p(x, x_1)}
\min
\{ I(X, X_1 ; Y),\;  H(Y_1 | X_1) + I(X; Y|X_1, Y_1) \}.
\label{eq:aref}
\end{equation}
Although this setup looks similar to ours, we note that neither
\eqref{eq:aref} nor Theorem~\ref{thm:main} implies the other.  In a
sense, our model is more deterministic in the relay-to-receiver link,
while the El Gamal--Aref model is more deterministic in the
transmitter-to-relay link.

A natural question arises whether our Theorem~\ref{thm:main} follows
from Theorem~\ref{thm:ceg} as a special case.  We first note that in
the coding scheme described in Section~\ref{sec:main}, the relay does
neither ``decode'' nor ``compress'', but instead ``hashes'' its received
output.  Indeed, as a coding scheme, this ``hash-and-forward'' appears
to be a novel method of summarizing the relay's information.  However,
``hash-and-forward'' is not the unique coding scheme achieving the
capacity
\[
C(R_0) = \max_{p(x)} \min \{ I(X;Y) + R_0,\; I(X;Y_1, Y) \}.
\]
In the next section, we show that ``compress-and-forward'' can achieve
the same rate.  

\section{Compress and Forward}
\label{sec:second}

Theorem~\ref{thm:main} was proved using ``hash-and-forward'' in
Section~\ref{sec:first}.  Here we argue that the capacity in
Theorem~\ref{thm:main} can also be achieved by
``compress-and-forward''.

We start with a special case of Theorem~\ref{thm:ceg}.  The
``compress-and-forward'' part (cf.\@
\cite[Theorem~6]{Cover--El-Gamal1979}), combined with the
relay-to-receiver communication of rate $R_0$, gives the achievable
rate
\begin{equation}
\label{eq:wz1}
R^*(R_0) = \sup I(X; Y, \hat{Y}_1),
\end{equation}
where the supremum is over all joint distributions of the form
$p(x) p(y,y_1|x) p(\hat{y}_1|y_1)$ satisfying
\begin{equation}
\label{eq:wz2}
I(Y_1;\hat{Y}_1|Y) \le R_0.
\end{equation}
Here the inequality \eqref{eq:wz2} comes from the Wyner--Ziv
compression~\cite{Wyner--Ziv1976} of the relay's output $Y_1^n$ based
on the side information $Y^n$.  The achievable rate \eqref{eq:wz1}
captures the idea of decoding $X^n$ based on the receiver's output
$Y^n$ and the compressed version $\hat{Y}_1^n$ of the relay's output
$Y_1^n$.

We now derive the achievability of the capacity
\[
C(R_0) = \max_{p(x)} \min\{ I(X; Y, Y_1), I(X;Y)+ R_0 \}
\]
from an algebraic reduction of the achievable rate given by \eqref{eq:wz1}
and \eqref{eq:wz2}.  First observe that, because of the deterministic
relationship $Y_1 = f(X, Y)$, we have
\[
I(X; \hat{Y}_1 | Y) \ge I(Y_1; \hat{Y}_1 | Y).
\]
Also note that, for any triple $(X,Y,Y_1)$, if $H(Y_1|Y) > R_0$, there
exists a distribution $p(\hat{y}_1|y_1)$ such that $(X,Y) \to Y_1 \to
\hat{Y}_1$ and $I(Y_1;\hat{Y}_1|Y) = R_0$.

Henceforth, maximums are taken over joint distributions of the form
$p(x)p(y,y_1|x)p(\hat{y}_1|y_1)$ with $Y_1 = f(X,Y)$.  We have
\begin{align}
R^*(R_0) 
&= \sup \{I(X; Y, \hat{Y}_1) : I(Y_1;\hat{Y}_1 | Y) \le R_0\} \notag\\
&\ge \sup \{ I(X; Y, \hat{Y}_1) : 
             I(Y_1;\hat{Y}_1 | Y) \le R_0, H(Y_1|Y) > R_0\}\notag\\
&\ge \sup \{ I(X; Y, \hat{Y}_1) : 
             I(Y_1;\hat{Y}_1 | Y) = R_0, H(Y_1|Y) > R_0\} \notag\\
&= \sup   \{ I(X; Y) + I(X; \hat{Y}_1 | Y) :
             I(Y_1;\hat{Y}_1 | Y) = R_0, H(Y_1|Y) > R_0 \} \notag\\
&\ge \sup \{ I(X; Y) + I(Y_1; \hat{Y}_1 | Y) :
I(Y_1;\hat{Y}_1 | Y) = R_0, H(Y_1|Y) > R_0 \} \notag\\
&= \sup \{ I(X; Y) + R_0 :
I(Y_1;\hat{Y}_1 | Y) = R_0, H(Y_1|Y) > R_0 \} \notag\\
&= \max \{I(X;Y) + R_0 :  H(Y_1|Y) > R_0 \}. \notag
\end{align}
On the other hand,
\begin{align}
R^*(R_0) 
&= \sup \{I(X; Y, \hat{Y}_1) : I(Y_1;\hat{Y}_1 | Y) \le R_0\} \notag \\
&\ge \sup \{ I(X; Y, \hat{Y}_1) : 
             I(Y_1;\hat{Y}_1 | Y) \le R_0, H(Y_1|Y) \le R_0\} \notag\\
&\ge \sup \{ I(X; Y, \hat{Y}_1) : 
             \hat{Y}_1 = Y_1, H(Y_1|Y) \le R_0\} \notag \\
&= \sup   \{ I(X; Y, Y_1) :
             \hat{Y}_1 = Y_1, H(Y_1|Y) \le R_0\} \notag\\
&= \max   \{ I(X; Y, Y_1) :
             H(Y_1|Y) \le R_0\}. \notag
\end{align}
Thus, we have
\begin{align*}
R^*(R_0) &\ge \max_{p(x): H(Y_1|Y) > R_0} I(X;Y) + R_0
\intertext{and}
R^*(R_0) &\ge \max_{p(x): H(Y_1|Y) \le R_0} I(X; Y, Y_1),
\intertext{and therefore,}
R^*(R_0) &\ge \max_{p(x)} \min\{ I(X;Y) + R_0,\; I(X; Y, Y_1) \}.
\end{align*}
In words, ``compress-and-forward'' achieves the capacity. 

\section{Discussion: Random Binning vs.\@ Random Covering}
\label{sec:discuss}

\begin{figure}[h]
\begin{center}
\input{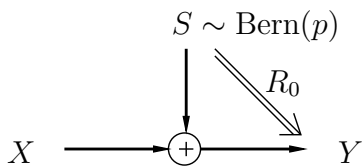}
\end{center}
\caption{Binary symmetric channel with rate-limited state information
at receiver.}
\label{fig:bsc}
\end{figure}

It is rather surprising that both ``hash-and-forward'' and
``compress-and-forward'' optimally convey the relay information to the
receiver, especially because of the dual nature of compression (random
covering) and hashing (random binning).  (And the hashing in
``hash-and-forward'' should be distinguished from the hashing in
Wyner--Ziv source coding.)  The example in Figure~\ref{fig:bsc}
illuminates the difference between the two coding schemes.

Here the binary input $X \in \{0,1\}$ is sent over a binary symmetric
channel with cross-over probability $p$, or equivalently, the channel
output $Y \in \{0,1\}$ is given as
\[
Y = X + S \pmod{2},
\]
where the binary additive noise $S \sim \textrm{Bern}(p)$ is
independent of the input $X$.  With no information on $S$ available at
the transmitter or the receiver, the capacity is
\[
C(0) = 1 - H(p).
\]

Now suppose there is an intermediate node which observes $S$ and
``relays'' that information to the decoder through a side channel of
rate $R_0$.  Since $S = X + Y$ is a deterministic function of $(X,Y)$,
Theorem~\ref{thm:main} applies and we have
\[
C(R_0) = 1 - H(p) + R_0
\]
for $0 \le R_0 \le H(p).$  

There are two ways of achieving the capacity.  First, hashing.  The
relay hashes the entire binary $\{0,1\}^n$ into $2^{nR_0}$ bins, then
sends the bin index $b(S^n)$ of $S^n$ to the decoder.  The decoder
checks whether a specific codeword $X^n(w)$ is typical with the
received output $Y^n$ and then whether $S^n(w) = X^n(w) + Y^n$ matches
the bin index.

\begin{figure}[h]
\begin{center}
\vspace*{12pt}
\input{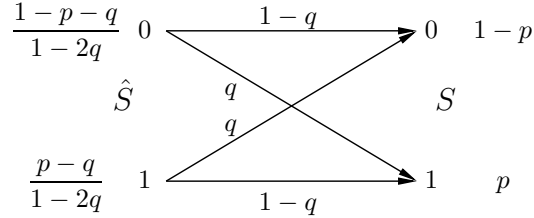}
\end{center}
\caption{Backward channel for the binary rate distortion problem.}
\label{fig:dist}
\end{figure}

Next, covering.  The relay compresses the state sequence $S^n$ using
the binary lossy source code with rate $R_0$.  More specifically, we
use the standard backward channel for the binary rate distortion
problem (see Figure~\ref{fig:dist}):
\[
S = \hat{S} + U. 
\]
Here $\hat{S} \in \{0,1\}$ is the reconstruction symbol
and $U \sim \textrm{Bern}(q)$ is independent of $\hat{S}$ (and $X$)
with parameter $q$ satisfying
\[
R_0 = I(S; \hat{S}) = H(p) - H(q).
\]  
Thus, using $n R_0$ bits, the ultimate receiver can reconstruct
$\hat{S}^n$.

Finally, decoding $X^n \sim \textrm{Bern}(1/2)$ based on $(Y^n,
\hat{S}^n)$, we can achieve the rate
\begin{align*}
I(X; Y, \hat{S}) &= I(X; X + S, S + U)\\
&\ge I(X; X + U) \\
&= 1 - H(q)\\
&= 1 - H(p) + R_0.
\end{align*}

In summary, the optimal relay can partition its received signal space
into either random bins or Hamming spheres.

The situation is somewhat reminiscent of that of lossless block source
coding.  Suppose $\{X_i\}$ is independent and identically distributed
(i.i.d.)  $\sim \textrm{Bern}(p)$.  Here are two basic methods of
compressing $X^n$ into $n H(p) $ bits with asymptotically negligible
error.

\begin{enumerate}
\item[1)] \emph{Hashing.} The encoder simply hashes $X^n$ into one of $2^{n
H(p)}$ indices.  With high probability, there is a unique
typical sequence with matching hash index.

\item[2)] \emph{Enumeration}~\cite{Cover1973}.  The encoder enumerates
$2^{n H(p)}$ typical sequences.  Then $n H(p)$ bits are required to
give the enumeration index of the observed typical sequence.  With
high probability, the given sequence $X^n$ is typical.
\end{enumerate}

While these two schemes are apparently unrelated, they are both
extreme cases of the following coding scheme.

\begin{enumerate}
\item[3)] \emph{Covering with hashing.} By fixing $p(\hat{x}|x)$ and
generating independent sequences $\hat{X}^n(i),$ $i = 1,\ldots,
2^{nI(X;\hat{X})},$ each i.i.d.\@ $\sim p(\hat{x})$, we can induce a
set of $2^{n I(X;\hat{X})}$ coverings for the space of typical
$X^n$'s.  For each cover $\hat{X}^n(i)$, there are $\approx 2^{n
H(X|\hat{X})}$ sequences that are jointly typical with $\hat{X}^n(i)$.
Therefore, by hashing $X^n$ into one of $2^{nH(X|\hat{X})}$ hash
indices and sending it along the cover index, we can recover a typical
$X^n$ with high probability.  This scheme requires $n(I(X;\hat{X}) +
H(X|\hat{X})) = nH(p)$ bits.
\end{enumerate}

Now if we take $\hat{X}$ independent of $X$, then we have the case of
hashing only.  On the other hand, if we take $\hat{X} = X$, then we
have enumeration only, in which case the covers are Hamming spheres of
radius zero.  It is interesting to note that the combination scheme
works under any $p(\hat{x}|x)$.

Thus motivated, we combine ``hash-and-forward'' with
``compress-and-forward'' in the next section.

\section{Compress, Hash, and Forward}
\label{sec:third}

Here we show that a combination of ``compress-and-forward'' and
``hash-and-forward'' can achieve
the capacity
\[
C(R_0) = \max_{p(x)} \min\{ I(X; Y, Y_1),\enspace I(X;Y)+ R_0 \}
\]
for the setup in Theorem~\ref{thm:main}.

We first fix an \emph{arbitrary} conditional distribution
$p(\hat{y}_1|y_1)$ and generate $2^{n(I(Y_1;\hat{Y}_1)+\epsilon)}$
sequences $\hat{Y}_1^n(i),$ $i =
1,2,\ldots,2^{n(I(Y_1;\hat{Y}_1)+\epsilon)},$ each i.i.d.\@ $\sim
p(\hat{y}_1)$.  Then, with high probability, a typical $Y_1^n$ has a
jointly typical cover $\hat{Y}_1^n(Y_1^n)$.  (If there is more than
one, pick the one with the smallest index.  If there is none, assign
$\hat{Y}_1^n(1)$.)

There are two cases to consider, depending on our choice of
$p(\hat{y}_1|y_1)$ (and the input codebook distribution $p(x)$).
First suppose
\begin{equation}
\label{eq:hat-r0}
I(Y_1; \hat{Y}_1|Y) \ge R_0.
\end{equation}
If we treat $\hat{Y}_1^n(Y_1^n)$ as the relay output, $\hat{Y}_1^n$ is
a deterministic function of $Y_1^n$ and thus of $(X^n, Y^n)$.
Therefore, we can use ``hash-and-forward'' on $\hat{Y}_1^n$ sequences.
(Markov lemma~\cite{Berger1978} justifies treating
$\hat{Y}_1^n(Y_1^n)$ as the output of the memoryless channel $p(y,
\hat{y}_1|x)$.)  This implies that we can achieve
\[
R^*(R_0) = \min\{ I(X; Y) + R_0, \enspace I(X; Y, \hat{Y}_1) \}.
\]
But from \eqref{eq:hat-r0} and the functional relationship between
$Y_1$ and $(X, Y)$, we have
\begin{align*}
I(X; Y, \hat{Y}_1) &= I(X; Y) + I(X; \hat{Y}_1|Y)\\
&\ge I(X; Y) + I(Y_1; \hat{Y}_1 | Y) \\
&\ge I(X; Y) + R_0.
\end{align*}
Therefore, 
\[
R^*(R_0) = I(X;Y) + R_0,
\]
which is achieved by the above ``compress-hash-and-forward'' scheme
with $p(x)$ and $p(\hat{y}_1|y_1)$ satisfying \eqref{eq:hat-r0}.

Alternatively, suppose 
\begin{equation}
\label{eq:hat-r0-2}
I(Y_1; \hat{Y}_1|Y) \le R_0.
\end{equation}
Then, we can easily achieve the rate $I(X; Y, \hat{Y}_1)$ by the
``compress-and-forward'' scheme.  The rate $R_0 \ge I(Y_1;\hat{Y}_1|Y)$
suffices to convey $\hat{Y}_1^n$ to the ultimate receiver.

But we can do better by using the remaining $\Delta = R_0 - I(Y_1;
\hat{Y}_1|Y)$ bits to further hash $Y_1^n$ itself.  (This hashing of
$Y_1^n$ should be distinguished from that of Wyner--Ziv coding which
bins $\hat{Y}_1^n$ codewords.)  By treating $(Y, \hat{Y}_1)$ as a new
ultimate receiver output and $Y_1$ as the relay output,
``hash-and-forward'' on top of ``compress-and-forward'' can achieve
\begin{equation}
R^*(R_0) = \min \{ I(X; Y, \hat{Y}_1) + \Delta, \enspace
I(X; Y, \hat{Y}_1, Y_1)\}.
\label{eq:r0-intermediate}
\end{equation}
Since
\begin{align*}
I(X; Y, \hat{Y}_1) + \Delta
&= I(X; Y, \hat{Y}_1) - I(Y_1; \hat{Y}_1|Y) + R_0 \\
&\ge I(X; Y, \hat{Y}_1) - I(X; \hat{Y}_1|Y) + R_0 \\
&= I(X; Y) + R_0
\end{align*}
and 
\[
I(X; Y, \hat{Y}_1, Y_1) = I(X; Y, Y_1),
\]
the achievable rate in \eqref{eq:r0-intermediate} reduces to
\[
R^*(R_0) = \min \{ I(X; Y) + R_0, \enspace I(X; Y, Y_1) \}.
\]
Thus, by maximizing over input distributions $p(x)$, we can achieve
the capacity for either case \eqref{eq:hat-r0} or
\eqref{eq:hat-r0-2}.

It should be stressed that our combined ``compress-hash-and-forward''
is optimal, regardless of the covering distribution
$p(\hat{y}_1|y_1)$.  In other words, any covering (geometric
partitioning) of $Y_1^n$ space achieves the capacity if properly
combined with hashing (nongeometric partitioning) of the same space.
In particular, taking $\hat{Y}_1 = Y_1$ leads to ``hash-and-forward''
while taking the optimal covering distribution $p^*(\hat{y}_1|y_1)$
for \eqref{eq:wz1} and \eqref{eq:wz2} in Section~\ref{sec:second}
leads to ``compress-and-forward''.

\section{Ahlswede--Han Conjecture}

In this section, we show that Theorem~\ref{thm:main} confirms the
following conjecture by Ahlswede and Han~\cite{Ahlswede--Han1983} on
the capacity of channels with rate-limited state information at the
receiver, for the special case in which the state is a deterministic
function of the channel input and the output.

\begin{figure}[h]
\begin{center}
\vspace*{12pt}
\input{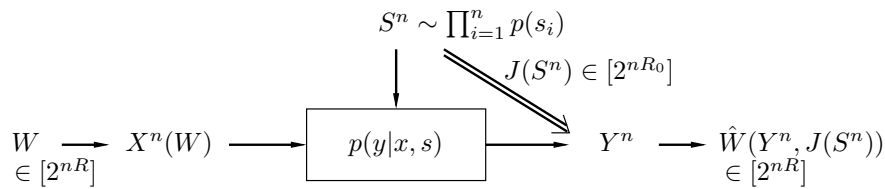}
\end{center}
\caption{Channel with rate-limited state information at the decoder.}
\label{fig:ah}
\end{figure}

First, we discuss the general setup considered by Ahlswede and Han, as
shown in Figure~\ref{fig:ah}.
Here we assume that the channel $p(y|x,s)$ has independent and
identically distributed state $S^n$ and the decoder can be informed
about the outcome of $S^n$ via a separate communication channel at a
fixed rate $R_0$.  Ahlswede and Han offered the following conjecture
on the capacity of this channel.

\begin{conjecture}[{Ahlswede--Han~\cite[Section V]{Ahlswede--Han1983}}]
The capacity of the state-dependent channel $p(y|x,s)$ as depicted in
Figure~\ref{fig:ah} with rate-limited state information available at
the receiver via a separate communication link of rate $R_0$ is given
by
\begin{equation}
\label{eq:ah}
C(R_0) = \max I(X; Y | \hat{S}),
\end{equation}
where the maximum is over all joint distributions of the form
$p(x)p(s) p(y|x,s) p(\hat{s}|s)$ such that
\[
I(S; \hat{S} | Y) \le R_0
\]
and 
the auxiliary random variable $\hat{S}$ has cardinality
$|\hat{\mathcal{S}}| \le |\mathcal{S}| + 1$.
\end{conjecture}

It is immediately seen that this problem is a special case of a relay
channel with a noiseless link (Figure~\ref{fig:det-relay}).  Indeed,
we can identify the relay output $Y_1$ with the channel state $S$ and
identify the relay channel $p(y, y_1|x) = p(y_1|x) p(y|x,y_1)$ with
the state-dependent channel $p(s) p(y|x,s)$.  Thus, the channel with
rate-limited state information at the receiver is a relay channel in
which the relay channel output $Y_1$ is independent of the input $X$.
The binary symmetric channel example in Section~\ref{sec:discuss}
corresponds to this setup.

Now when the channel state $S$ is a deterministic function of $(X,
Y)$, for example, $S = X + Y$ as in the binary example in
Section~\ref{sec:discuss}, Theorem~\ref{thm:main} proves the
following.

\begin{theorem}
For the state-dependent channel $p(y|x,s)$ with state information
available at the decoder via a separate communication link of rate
$R_0$, if the state $S$ is a deterministic function of the channel
input $X$ and the channel output $Y$, then the capacity is given by
\begin{equation}
\label{eq:cap-ah}
C(R_0) = \max_{p(x)} \min\{I(X;Y) + R_0, \; I(X;Y,S)\}.
\end{equation}
\end{theorem}

Our analysis of ``compress-and-forward'' coding scheme in
Section~\ref{sec:second} shows that \eqref{eq:ah} reduces to
\eqref{eq:cap-ah}, confirming the Ahlswede--Han conjecture when $S$ is
a function of $(X,Y)$.  On the other hand, our proof of achievability
(Section~\ref{sec:first}) shows that ``hash-and-forward'' is equally
efficient for informing the decoder of the state information.

\section{Concluding Remarks}
Even a completely oblivious relay can boost the capacity to the cut
set bound, if the relay reception is fully recoverable from the
channel input and the ultimate receiver output.  And there are two
basic alternatives for the optimal relay function---one can either
compress the relay information as in the traditional method of
``compress-and-forward,'' or simply hash the relay information.  In
fact, infinitely many relaying schemes that combine hashing and
compression can achieve the capacity.  While this development depends
heavily on the deterministic nature of the channel, it reveals an
interesting role of hashing in communication.

\def\cprime{$'$}

\end{document}